\documentclass[lettersize,journal]{IEEEtran}
\usepackage{algorithmic}
\usepackage{array}
\usepackage{textcomp}
\usepackage{stfloats}
\usepackage{url}
\usepackage{verbatim}
\usepackage{graphicx} 	
\usepackage{cite}

\usepackage{amsmath,amssymb,amsfonts}
\usepackage{url}

\usepackage{subcaption}
\usepackage[export]{adjustbox}
\usepackage{capt-of}
\usepackage{caption}
\usepackage{booktabs}
\usepackage{float}
\usepackage{multirow}
\usepackage[table,xcdraw]{xcolor}
\usepackage{multirow}
\usepackage{enumitem}
\usepackage[export]{adjustbox}
\usepackage[linesnumbered,ruled,vlined]{algorithm2e}
\usepackage{eurosym}
\makeatletter
\renewcommand{\@IEEEsectpunct}{~}
\makeatother

\newcommand{\squeezeup}{\vspace{-1.5mm}}

\begin{document}
\bstctlcite{IEEEexample:BSTcontrol}
\title{How to build a sovereign network? - A proposal to measure network sovereignty}

\author{Shakthivelu Janardhanan,~\IEEEmembership{Student Member,~IEEE,}
Ritanshi Agarwal,
Wolfgang Kellerer,~\IEEEmembership{Senior Member,~IEEE,}
Carmen Mas-Machuca,~\IEEEmembership{Senior Member,~IEEE.}

% \squeezeup \squeezeup \squeezeup \squeezeup \squeezeup \squeezeup

        % <-this % stops a space
\thanks{This work has received funding from the Bavarian Ministry of Economic Affairs, Regional Development, and Energy under the project "6G Future Lab Bavaria" and by the Federal Ministry of Education and Research of Germany in the program of “Souver\"an. Digital. Vernetzt.”. Joint project 6G-life- 16KISK002.}% <-this % stops a space
%\thanks{Manuscript received April 19, 2021; revised August 16, 2021.}}
}
\markboth{Journal of \LaTeX\ Class Files,~Vol.~14, No.~8, August~2021}%
{Shell \MakeLowercase{\textit{et al.}}: A Sample Article Using IEEEtran.cls for IEEE Journals}


\maketitle

\begin{abstract}

Network sovereignty is a network operator's ability to reduce the dependency on component manufacturers to minimize the impact of manufacturer failures.
Network operators now face new design challenges to increase network sovereignty and avoid vendor lock-in problems because a high dependency on a manufacturer corresponds to low survivability if that manufacturer is unavailable.
% We aim to provide guidelines to assign manufacturers to the nodes in the topology such that network sovereignty is maximized.
The main contribution of this work is the proposal of a novel metric to measure network sovereignty, the Cut Set Coloring (CSC) score. Based on the CSC core metric \textit{CSC-ILP}, our Integer Linear Program formulation is presented to maximize network sovereignty. 
We compare \textit{CSC-ILP}'s performance with state of the art manufacturer assignment strategies.

\end{abstract}

\section{Introduction}
\label{chap:intro}
%ADd country examples  
%JUniper JUNOS example
%2021, OVHcloud, Europe’s largest cloud-hosting

Though network operators can buy components from different manufacturers, they favor only one or two manufacturers to ease interoperability while saving costs on bulk purchasing and deployment.
However, this strategy causes a high dependency on them. 
If one of those manufacturers becomes unavailable due to inherent component design flaws, geopolitical ban, subcomponent manufacturers becoming unavailable, security vulnerabilities, etc., all components of that manufacturer are compromised simultaneously. 
We define a \textit{manufacturer failure} as the simultaneous unavailability of all components from that manufacturer due to any of the aforementioned reasons.
%Such massive correlated simultaneous failures must be avoided. 
Operators aim to reduce the impact of such massive correlated simultaneous failures.
%Many such manufacturer failures have occurred recently. 
%For example, Samsung Galaxy 7 units exploded~\cite{samsung} in 2016 due to defective battery. 
%A software issue can also cause multiple failures. 
%%For example, the Equifax data breach~\cite{equifax} in 2017 was due to a security vulnerability in an outdated Apache Struts library. 
%For example, a vulnerability in Log4j, an industrial-grade logging framework~\cite{canada}, shut down 4000 Canadian government websites in 2021. %Both these software issues led to the leaking of critical user data. 
%Such incidents invariably led to large downtimes and financial losses.

Several examples of different types of such manufacturer failures have been recorded.
%Such manufacturer failures can be caused by several factors. 
It can be a natural disaster or an accident like the fire that burned down the OVHcloud data centers in 2021. An estimated 3.6 million websites and 464,000 domain names were affected~\cite{dcn-fire}, causing an estimated loss of \euro 105 million. 
From 2014 to 2023,~\cite{dcn-fire} has identified 22 major data center fires.
Such accidents furthered the cause for backup websites and data storage hosted on different cloud platforms to avoid a singular dependency. % on any cloud host.
Such failures can also be inadvertently man-made. The recent CrowdStrike outage~\cite{crowdstrike} of 2024 caused an estimated 8.5 million devices to fail simultaneously, resulting in losses of over \$5.4 billion in revenue. This was caused by a singular dependency on CrowdStrike and Windows. 
Such failures can also be security vulnerabilities, as recorded in the CVE vulnerabilities list~\cite{cvelist}. For instance, CVE-2021-0275, a cross-site scripting vulnerability, affected Juniper Networks Junos Operating System (OS)~\cite{cvelist}, which is a popular OS for enterprise and Data Center Networks (DCNs). Similarly, CVE-2013-6026 was a firmware backdoor that allowed attackers to remotely access several D-Link, Alpha Networks, and Planex routers~\cite{cvelist}. Such security issues form a shared vulnerability because all the routers from a vendor use the same firmware.
%To combat such dangerous dependencies, researches suggest diversifying the network and storage facilities with heterogeneous vendors to avoid shared vulnerabilities~\cite{tipper, amir, ghawash, mine-Naga}.

%Another software issue, caused by a vulnerability in Log4j, an industrial-grade logging framework~\cite{canada}, shut down 4000 Canadian government websites in 2021. These dependencies have been identified critical by several nations and questions have been raised on the national data localization policies. The urgent need to avoid governmental dependency on foreign companies also points towards digital sovereignty and autonomy~\cite{tipper}.

%\sha{I can also write about the pager and walkietalkie attacks recently in Lebanon- but this is active war stuff, i do not want to comment on it.}

Recently, governments are removing foreign dependencies from core infrastructure. For example, the United States, Spain, England, Germany, etc. have banned Chinese equipment from their core networks~\cite{tipper, mine-Naga, huawei_US_1}. 
%Other nations like Spain, England, Germany, etc. also followed suit.
~\cite{li_interpret} details the policy-making trends on digital autonomy in several countries, like Russia's sovereign network (RuNet) and financial services via Mir~\cite{tipper} to the European Union's GAIA-X initiative, to reduce the dependency on American cloud services. These policies aim to remove foreign dependencies and vendor-related threats in the critical infrastructure.

% ~\cite{Avizienis:2004}
%Availability, reliability, safety, integrity, and maintainability are traditional dependability attributes~\cite{Avizienis:2004} that can't quantify manufacturer-related simultaneous correlated failures. 
%Traditional dependability attributes~\cite{Avizienis:2004} like availability can not quantify manufacturer-related simultaneous correlated failures. 
%To understand why, let us differentiate network sovereignty from availability with an example of a s.   

% In this work, a flow refers to a unit of traffic between a source and destination.
A system's availability is the probability that it delivers uninterrupted service at any instant~\cite{Avizienis:2004}. Extending this definition, a network's availability is the probability that it can route all its flows at any instant. In this work, a flow refers to a unit of traffic between a source and destination. A networking device's availability is provided based on stress tests on the device in its datasheet through parameters such as mean time to fail (MTTF), mean time to repair (MTTR), etc. The availabilities of all the devices in the network can be used to derive a network's availability using tools like the Reliability Block Diagram (RBD). 

\begin{center}
\begin{figure*}[t]
\includegraphics[width=0.8\linewidth, center]{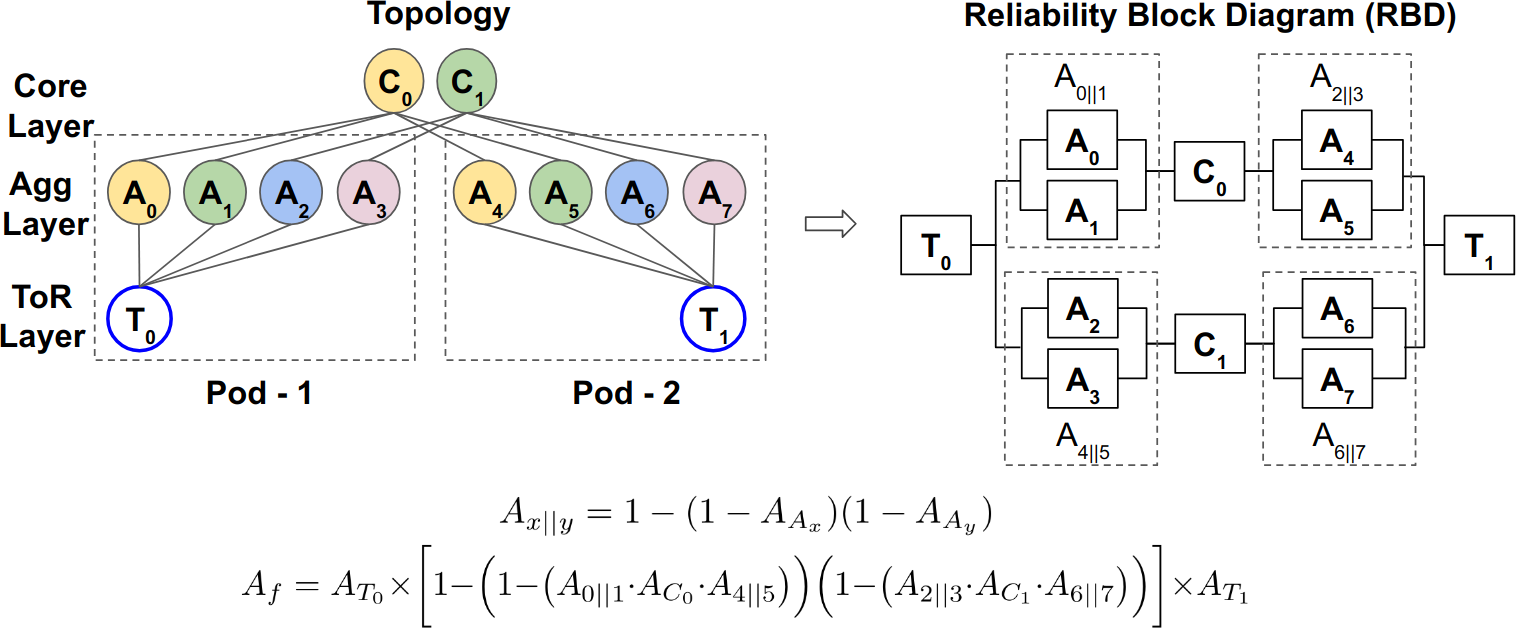} 
\caption{Network sovereignty vs. availability example, using a DCN topology and its RBD. $A_f$ and $A_n$ denote the flow and node availabilities, respectively. $A_{x||y}$ denotes the availability of two nodes $x$ and $y$ in parallel. Each node color represents a different manufacturer.} %The coloring of the nodes is explained in Section~\ref{sec:example-applicaiton}}
\label{fig:1}
\end{figure*}
\end{center}

Network availability is often confused with sovereignty. To understand the difference, let us consider the following example with a simplified DCN topology with twelve nodes, as shown in Fig.~\ref{fig:1}.

Let us first compare the availability and sovereignty for the DCN when all twelve nodes are from a Manufacturer-A ($M_A$) with node availabilities ($A_n$) of 0.999. 
Let there be a flow between nodes $T_0$ and $T_1$. 
This flow's availability ($A_f$) can be evaluated using the RBD in Fig.~\ref{fig:1}.
%The RBD in Fig.~\ref{fig:1} is used to evaluate the flow availability . 
When elements in an RBD are in series, their availabilities are multiplied, while parallel elements are combined as the probability that at least one of them is working. In Fig.~\ref{fig:1}, (i)~$A_0$, $A_1$, $A_2$, and $A_3$ are in parallel, (ii)~$C_0$ and $C_1$ are in parallel, (iii)~$A_4$, $A_5$, $A_6$, and $A_7$ are in parallel, and (iv)~the subsystems in (i), (ii), and (iii) are in series with the source ($T_0$) and destination ($T_1$) nodes. 
Using the equation for $A_f$ in Fig.~\ref{fig:1}, $A_f$ is calculated as 0.998. However, if $M_A$ fails, all the nodes in the network fail, and the flow also fails due to the overwhelming dependency on $M_A$. Hence, this case has minimal sovereignty.

\vspace{-1cm}
Let us consider the second case, where the nodes are from four different manufacturers- $M_A$ (yellow), $M_B$ (green), $M_C$ (blue), and $M_D$ (pink) with $A_n$ of 0.999, 0.995, 0.990, and 0.985, respectively. Let $T_0$ and $T_1$ be from $M_C$ and $M_D$, respectively.
Now, with the same RBD but different $A_n$, $A_f$ drops to 0.975. However, if one manufacturer (for example, $M_A$) fails, several working paths remain between $T_0$ and $T_1$. 
Therefore, the dependency on $M_A$ is avoided by introducing more manufacturers.
This case shows higher sovereignty but lower availability than the first example.

%The first case has higher availability with lower sovereignty. The second case has lower availability but higher sovereignty.

The above example shows that dependability attributes like availability cannot represent the impact of manufacturer failures.
To address this shortcoming, technology sovereignty~\cite{edler2020technology} was defined as an organization's ability to provide a technology without any dependencies on suppliers. 
Extending to communication networks, we define network sovereignty as a network operator's ability to operate a network without any dependencies on component manufacturers. %Other sovereignty-related works are discussed in Section~\ref{chap:bg}.
% The absence of a metric to measure network sovereignty is a critical challenge and the focus of this paper. 

Given a network topology, an operator must decide how many manufacturers to buy nodes from and where to place these nodes. 
%From the examples in Sections~\ref{chap:intro} and~\ref{chap:metric}, 
%An appropriate placement of nodes from different manufacturers is vital to building a sovereign network. 
The research question in this work is the Manufacturer Assignment for Sovereignty (MAS) problem~\cite{mine-Naga, mine-ICTON-demo}:
\begin{enumerate}[label=(\Alph*), start=17]
	\item\label{q:Q1} Given a network topology as a set of interconnected nodes and the number of node manufacturers, what is the best manufacturer assignment possible to maximize network sovereignty? In other words, how many nodes should be purchased from each manufacturer, and where must they be placed in the topology?
\end{enumerate}
The contributions of this work to that question are as follows.
\begin{enumerate}
\item Introducing a novel network sovereignty metric- `Cut Set Coloring (CSC) score' (Section~\ref{chap:metric}).
\item Evaluating the most sovereign manufacturer assignment with our CSC score-based ILP formulation, \textit{CSC-ILP} to solve the MAS problem (Section~\ref{chap:CSC-ILP}).
\item Comparing \textit{CSC-ILP}'s performance with centrality metrics-based heuristics and state of the art works like \textit{Diversity Assignment Problem (DAP)}~\cite{amir}, \textit{Min cut set heuristic}~\cite{mine-ICTON-demo} and \textit{Naga}~\cite{mine-Naga} (Section~\ref{chap:evalworkflow}).
\item Deriving design guidelines for network sovereignty (Section~\ref{chap:results}).
\end{enumerate}

%For the first case, let all six nodes be from $M_A$ with node availabilities of 0.999. Let there be a flow between nodes $S$ and $T$. Using the Reliability Block Diagram (RBD) in Fig.~\ref{fig:1} for the flow, the flow availability can be calculated as 0.998. However, if $M_A$ fails, all the nodes in the network fail, and the flow fails due to the overwhelming dependency on $M_A$.

%Let us consider the second case, where the nodes are bought from two different manufacturers. Let nodes $A$, $B$, and $C$ be from $M_A$ with node availabilities 0.999 while the nodes $S$, $C$, $D$, and $T$ are from Manufacturer-B with node availabilities 0.995. In this case, the flow availability, calculated from the RBD in Fig.~\ref{fig:1}, drops to 0.98. Now, if $M_A$ fails, only the nodes $A$, $B$, and $C$ fail. However, $S$-$D$-$E$-$T$ is still a working path between $S$ and $T$. This case effectively avoids the dependency on $M_A$ by introducing one more manufacturer. 

%\begin{equation*}
%A_f = A_{T_0} \times \bigl(1-\prod\limits_{i=0}^{3}(1 - A_{A_i})\bigr) \times \bigl(1-\prod\limits_{i=0}^{1}(1 - A_{C_i})\bigr) \times \bigl(1-\prod\limits_{i=4}^{7}(1 - A_{A_i})\bigr) \times A_{T_1}
%\end{equation*}

\section{Sovereignty and multi-vendor networks in literature}
\label{chap:bg}

%The term sovereignty was first documented in the fields of politics and economics. However, d
%Digital and technology sovereignty became popular with the digital revolution in the late $20^{th}$ century. 
%In the $21^{st}$ century, the Great Firewall of China, the national internet of Iran, Brazil's plan to lay submarine cables to connect to Europe, Germany's plan to exclude foreign contracts in its national cloud services, and Europe's EU cloud or the Schengen cloud concept serve as striking examples of every region trying to maintain its own digital autonomy~\cite{broeders2016public}.

Recent trade wars, sanctions, and economic crises highlight today's need for technology, data, and network sovereignty.% and, therewith, data and network sovereignty is at an all-time high.
Network sovereignty differs from network redundancy and diversity~\cite{mine-Naga, tipper}. Redundancy by duplicating network elements to increase network paths may improve availability. However, sovereignty is not improved if all components are from the same manufacturer. In contrast, network diversity provides multiple technology and/or operator-independent paths to connect to the network. However, if the same operator provides both technologies and/or the operators use components from the same manufacturer, sovereignty is not improved. Therefore, network sovereignty aims to provide multiple technology, operator, and manufacturer-independent paths in the network. %between any source and destination in the network. %However, there will be limitations on the number of technologies, operators, and manufacturers available in real-life networks.

%The current literature on multi-vendor networks focuses on the modeling~\cite{vickybro1}, interoperability issues~\cite{multi_vendor}, and
%network availability improvement~\cite{mine4}, 
%network security and trust guarantees~\cite{trust}, and techno-economic analysis~\cite{bauschert_vendor_selection}. 
%These works do not study the impact of manufacturer failures on network performance.

%The work in~\cite{mine2} focuses on the impact of multiple manufacturers in DCNs and provides guidelines to DCN operators to improve sovereignty.
%~\cite{mine2} tested heuristic manufacturer assignments and compared their performances.
%However,~\cite{mine2} has two major limitations- it did not quantitatively measure a DCN's sovereignty and is suitable only for symmetric DCN topologies.

Common issues in establishing sovereignty are geopolitics, multi-layer-multi-domain security, and lack of quantification metrics~\cite{mine-Naga, tipper, amir, mine-ICTON-demo, ghawash}.~\cite{amir} deals with the \textit{Diversity Assignment Problem (DAP)} for improved security and resilience. \textit{DAP}-ILP aims to introduce diversity among homogeneous routing nodes by introducing `variant nodes' to avoid shared vulnerabilities between the nodes. Since DAP aims to remove dependencies on particular node configurations,~\cite{amir} can be viewed as a work on network sovereignty and compared against our \textit{CSC-ILP}. Similarly,~\cite{ghawash} also employs node heterogeneity through the \textit{(r, s)-robustness with coloring} method to improve resilience against adversaries. However,~\cite{amir, ghawash} do not scale well with the network size.% Hence,~\cite{amir} proposed heuristics instead. 
%We compare our  to \textit{DAP}~\cite{amir} in Section~\ref{chap:results}.

Our previous work~\cite{mine-ICTON-demo} on network sovereignty proposes a heuristic manufacturer assignment, called the \textit{Min cut set heuristic}, based on minimal cut sets (MCSs). 
A cut set of a flow in a network is a set of elements that, when failed, causes the flow to fail.
A cut set is minimal if it can not be reduced further while still being a cut set. 
For example, in the topology in Fig.~\ref{fig:1}, $(A_0, A_1, A_2, A_3)$ is a MCS. %$(A_0, A_1, A_2, A_3, C_0, A_6)$ is a cut set while 
%Though the source and destination of a flow are the flow's MCSs, they are discarded because if the source or destination fails, the flow fails~\cite{amir, mine-Naga, mine-ICTON-demo}.
The source and destination are discarded as MCSs because if they fail, the flow fails~\cite{amir, mine-Naga, mine-ICTON-demo}.
The \textit{Min cut set heuristic} solves the MAS problem by maximizing the number of manufacturers (manufacturer diversity) inside the MCSs for each flow in the network to avoid a single point of failure. However,~\cite{amir, ghawash, mine-ICTON-demo} can not quantify a network's sovereignty. %Moreover,~\cite{mine-ICTON-demo} presents a heuristic, which is most likely to provide a sub-optimal manufacturer assignment.

We proposed the Path Set Diversity (PSD) score metric~\cite{mine-Naga} to measure and compare the network sovereignty of different manufacturer assignments. \textit{Naga}, an ILP formulation, was also proposed to solve the MAS problem. 
%The PSD score metric rewards each path based on the degree of manufacturer diversity in its paths. 
The PSD score for each path of a flow is equal to the reciprocal of the number of the number of manufacturers in that path. This is because, several manufacturers in a path denotes a dependence on all of those manufacturers for the flow to be successful.
For example, in Fig.~\ref{fig:1} with four manufacturers, the path-PSD score for the path $T_0$-$A_0$-$C_0$-$A_4$-$T_1$ is 1 because only yellow nodes are present in the path. The path-PSD score for the path $T_0$-$A_1$-$C_0$-$A_4$-$T_1$ is $\frac{1}{2}$ because there are two manufacturers in the path (green and yellow). 
The flow-PSD score is equal to the sum of all the path-PSD scores. However, paths with the same manufacturer combination should be counted only once. For example, the paths $T_0$-$A_2$-$C_1$-$A_7$-$T_1$ and $T_0$-$A_3$-$C_1$-$A_6$-$T_1$ have the same manufacturers in the path (blue, green, and pink). The order and the number of nodes in the path are irrelevant for a sovereignty study. The network-PSD score is the average of all the flow-PSD scores. \textit{Naga}~\cite{mine-Naga} maximizes the network-PSD score to solve the MAS problem.

In Fig.~\ref{fig:1}, there are eight simple paths between $T_0$ and $T_1$. However, the number of simple paths for each flow in a network can be tens of thousands. Additionally, there are several flows in a network adding to the problem complexity. Therefore, \textit{Naga} considers only $k$ shortest paths, where choosing an appropriate $k$ is a challenge. 
\textit{Naga} does not scale for large networks and/or a large $k$.
Furthermore, choosing the $k$ shortest paths for a symmetric topology with comparable link lengths, like the DCN topology in Fig.~\ref{fig:1}, is ambiguous. If the paths are not prioritized based on link lengths, the PSD score may not quantify the network's sovereignty accurately and \textit{Naga} may not produce the optimal manufacturer assignment.

%Now, our primary goal is to address the issues in previous works. Hence, 
In this work, we propose the novel CSC score metric to measure and compare the network sovereignty of manufacturer assignments while ensuring scalability and application on asymmetric and symmetric topologies. %Additionally, we propose \textit{CSC-ILP}, an ILP formulation to obtain the most sovereign network manufacturer assignment. We also compare the manufacturer assignments from \textit{CSC-ILP} with those based on (i)~centrality metrics, (ii)~minimal cut set heuristic~\cite{mine-ICTON-demo}, and (iii)~\textit{Naga}~\cite{mine-Naga}.

%Challenges
\section{Cut Set Coloring (CSC) score as a network sovereignty metric}
\label{chap:metric}
% This section discusses the proposed network sovereignty metric- the Path Set Diversity (PSD) score. First, we introduce the guidelines for developing a sovereign network. Then, we define our metric. he MCSs are , $(A, C, E)$, $(B, D)$, and $(B, C, E)$.
% This section discusses the Cut Set Coloring (CSC) score, our novel metric to quantify network sovereignty,  based on the minimal cut sets (MCSs) in the network. 
% A cut set of a flow in a network is a set of elements when failed, causes the flow to fail.
% A cut set is minimal if it can not be reduced any further while still being a cut set. 
% For example, in the example topology in Fig.~\ref{fig:1}, $(A_0, A_1, A_2, A_3, C_0, A_6)$ is a cut set while $(A_0, A_1, A_2, A_3)$ is an MCS.
% Though the source and destination of a flow are the flow's MCSs, for our sovereignty study, they are not considered because if the source or destination fails, the flow always fails.

\subsection{Towards a sovereign network}
\label{sec:guideline_for_sov}
%By the definition of a flow's MCS, a
A minimal cut set (MCS) gives the set of components that, when failing, causes the flow to fail. 
%In other words, all the nodes in a MCS must fail for that flow to fail. 
If all the MCS's nodes are from the same manufacturer, and if that manufacturer fails, the MCS fails, causing the flow to fail. Therefore, maximizing manufacturer diversity within an MCS is necessary to avoid a dependency on one manufacturer. Hence, the main guideline is:
%\cmas{why to use M1 as there is one guideline? and aren't the guidelines listed in Section 6D?}
%\begin{enumerate}[label=(\roman*)]
%\item Each MCS must have nodes from as many different manufacturers as possible to prevent the simultaneous failure of all the nodes in the MCS.
%\end{enumerate}
\textit{Each MCS must have nodes from as many different manufacturers as possible to prevent the simultaneous failure of all the nodes in the MCS.}

For example, consider the MCS $(A_0, A_1, A_2, A_3)$ in the topology in Fig.~\ref{fig:1}. If the MCS's nodes are from the same manufacturer (first case in Section~\ref{chap:intro}), the MCS fails if that manufacturer is unavailable, causing flow failure.
However, if the MCS's nodes are from different manufacturers (second case in Section~\ref{chap:intro}), then the MCS fails only if all the manufacturers in the MCS fail simultaneously. However, the simultaneous failure of multiple manufacturers is unlikely. Hence, such a diversified manufacturer assignment improves sovereignty by removing the dependency on manufacturers.
The aforementioned guideline must be followed for every MCS of each flow in the network. 

%From the above example, having multiple manufacturers and the appropriate manufacturer assignment improves network sovereignty. When the manufacturers are assumed to be colors, this problem requires the operator to color the MCSs appropriately to increase the diversity inside an MCS. Though this resembles a graph-coloring problem, it is not one because the number of manufacturers (colors) is limited. 

% TO do Shakthi, remove symbols as much as possible. all omegas
\subsection{Cut Set Coloring (CSC) score}
\label{sec:csc-score-description}
%Based on the guideline in Section~\ref{sec:guideline_for_sov}, we define our CSC score metric.
The MCS's CSC score, called the MCS-CSC score, is qualitatively defined as the extent of manufacturer diversity in that MCS. 
%Let the set of manufacturers be $M$, and each manufacturer is denoted by $m \in M$. Let $q_{mrj}$ be the binary variable equal to 1 if flow $r$’s $j^{th}$ MCS uses manufacturer $m$. $\kappa_{rj}$ is the $j^{th}$ MCS of flow $r$ and $K_r$ is the set of all MCSs of flow $r$. Mathematically, the CSC score ($\theta_{\text{r,j}}$) of the $j^{th}$ MCS of flow $r$ is given by,
% \begin{equation}
% \label{eq:1}
% \theta_{r,j} = 
% \begin{cases}7
%     0, & \text{if} \sum\limits_{m \in M}q_{mrj} = 1 \\ % That is, only one manufacturer used in the jth cutset
%     \frac{\sum\limits_{m \in M}q_{mrj}}{|\kappa_{rj}|} , & \text{otherwise}
% \end{cases} 
% \end{equation}
%The MCS-CSC score depends on the number of manufacturers in the MCS. 
If there is only one manufacturer, the MCS-CSC score should be penalized with a score of 0. If there is more than one manufacturer in the MCS, the singular dependency is avoided, and this is rewarded by a score equal to the number of manufacturers in the $i^{\text{th}}$ MCS ($|M^i|$) divided by the number of nodes in the MCS ($|n^i|$). If all the MCS's nodes are from different manufacturers, then the MCS has the highest MCS-CSC score of 1. Quantitatively,
\begin{equation}
 \label{eq:mcscsc}
 \text{MCS-CSC score} = 
 \begin{cases}
     0, & \text{if} |M^i| = 1 \\
     \frac{|M^i|}{|n^i|}, & \text{otherwise}
 \end{cases} .
\end{equation}
%The flow's CSC score called the Flow-CSC score, is the average of the MCS-CSC scores of all the MCSs of that flow. Similarly, the Network-CSC score is the average of the Flow-CSC scores of all the flows in the network. This Network-CSC score enables us to compare the sovereignty of different manufacturer assignments and networks. 
A flow is only as sovereign as its weakest MCS~\cite{tipper}. Therefore, the flow's CSC score called the Flow-CSC score, is equal to the least MCS-CSC scores of all the MCSs of that flow. %This idea also resonates with the proposal in~\cite{tipper}.
The Network-CSC score combines all the Flow-CSC scores in the network, represented as a range in a box plot. It can not be denoted by a single value, for example, as an average of all the Flow-CSC scores, because information on the weakest Flow-CSC score and the variance of the Flow-CSC scores might get lost. 
For example, two assignments may have the same mean values, but one assignment may have a worse weakest Flow-CSC score and must be avoided.
%The Network-CSC score enables comparing the sovereignty of different manufacturer assignments and networks. 

\subsection{Application of the CSC score}
\label{sec:example-applicaiton}
Consider the example DCN topology in Fig.~\ref{fig:1}. 
Consider four manufacturers available to the operator. 
Consider a flow between $T_0$ and $T_1$. 
The MCSs for this flow are $\{ (A_0, A_1, A_2, A_3), (A_0, A_1, C_1), (A_0, A_1, A_6, A_7), (C_0, C_1)$, $(C_0, A_2, A_3), (C_0, A_6, A_7), (A_4, A_5, C_1), (A_4, A_5, A_6, A_7)\}$. 
For the MCSs $(A_0, A_1, C_1)$ and $(A_4, A_5, C_1)$, only two manufacturers (yellow and green) are present for three nodes. Hence, from Eq.~\ref{eq:mcscsc}, they have MCS-CSC scores of $\frac{2}{3} = 0.66$ each. For each of the other MCSs, the number of manufacturers ($|M^i|$) equals the number of nodes ($|n^i|$). Therefore, they all have a score of 1 each.
For this flow, the minimum MCS-CSC score is 0.66. Therefore, the Flow-CSC score is also 0.66. For a network, the Network-CSC score is represented as a range of Flow-CSC scores calculated for each flow.

\subsection{Problem formulation: CSC-ILP}
\label{chap:CSC-ILP}

Consider a network topology, with vertices $V$ and edges $E$.
Let each flow be defined by its source and destination. 
We remove all the flows with a one-hop path because they do not have any other MCSs.
Let the available manufacturer list be $M$. Each flow has its set of MCSs. The order of the MCSs is irrelevant.

\textit{CSC-ILP} solves the MAS problem in~\ref{q:Q1} by assigning each node to one manufacturer such that the network sovereignty is maximized. For this purpose, \textit{CSC-ILP} identifies the minimum MCS-CSC score among all the flows in the network. Then, this MCS-CSC score is maximized by increasing the number of manufacturers in that MCS. This global optimization ensures the improvement in all the flows and, thereby, the network's sovereignty. 

As a secondary objective to improve the results, the sum of all the flows' MCS-CSC scores is maximized. 
This maximizes the diversity in as many MCSs as possible.
%This ensures that even if one MCS has no diversity, causing the Flow-CSC score to be zero, the other MCSs of the flow have more diversity.

For a network operator, the most important parameter is cost. 
Each manufacturer's node has a different cost. 
Therefore, \textit{CSC-ILP} considers the operator's cost constraint. The cost assumptions in this work are discussed in Section~\ref{sec:expsetup}.

\section{Case studies}
\label{chap:evalworkflow}

\subsection{Experimental setup}
\label{sec:expsetup}
%To analyze the performance of our CSC score metric and \textit{CSC-ILP}, we consider three core networks: (i)~Abilene~\cite{sndlib} (United States of America)- 11 vertices (one node with nodal degree of one has been removed) and 14 edges, (ii)~Polska~\cite{sndlib} (Poland)- 12 vertices and 18 edges, and (iii)~Germany\_17~\cite{sndlib} (Germany)- 17 vertices and 26 edges. Any-to-any traffic is considered.
We evaluate the CSC score metric and \textit{CSC-ILP} on four core networks~\cite{sndlib}: (i)~Abilene ($|V|=11$ (one node with nodal degree of one is removed), $|E|=14$), (ii)~Polska ($|V|=12$, $|E|=18$), (iii)~Germany\_17 ($|V|=17$, $|E|=26$), and (iv)~Nobel\_EU ($|V|=28$, $|E|=41$). We consider any-to-any traffic and four different numbers of manufacturers ($2,3,4,5$) in the market to compare varying degrees of competitive markets. 

To include the operator's cost constraint as per Section~\ref{chap:CSC-ILP}, in this work, we assume that the different manufacturers' node costs ($C_m$) vary linearly between 0.995 and 0.985. For example, if there are two manufacturers, then $C_0$ and $C_1$ are 0.995 and 0.985, respectively. If there are three manufacturers, then $C_0$, $C_1$, and $C_2$ are 0.995, 0.99, and 0.985, respectively.
Since there is no operator cost budget data available, we assume that the cost threshold ($C_T$) is equal to the product of the number of nodes in the network ($|V|$) and the mean possible cost per node, i.e., $|V| \times \frac{0.995 + 0.985}{2}$. 
$C_m$ and $C_T$ are \textit{CSC-ILP}'s input parameters that the operator must update as per their data. 

Though we evaluate core networks, the CSC score metric and \textit{CSC-ILP} apply to any network.

\subsection{Physical setup and run time}
\textit{CSC-ILP} uses Python-Gurobi optimization suite~\cite{gurobi} and was run on an AMD Ryzen 3700X octa-core processor with 32GB RAM.
\textit{CSC-ILP}'s run time for Nobel\_EU for five manufacturers was around 2 minutes, showing strong scope for scalability. 
On the same physical setup, \textit{DAP} took five hours for Nobel\_EU for four manufacturers, and did not deliver a result for five manufacturers after eight hours, highlighting the scalability issue discussed in~\cite{amir}.
%However, the run time of the \textit{CSC-ILP} is immaterial to our network planning problem.

\subsection{State of the art to compare with \textit{CSC-ILP}}
%Since no previous works solve the MAS problem in~\ref{q:Q1}, we evaluate manufacturer assignments based on centrality metrics-based heuristics to compare our \textit{CSC-ILP}'s performance. 
Different solutions have been evaluated and compared with \textit{CSC-ILP}. 

Centrality metrics evaluate a node's connectivity characteristics. Higher centrality means the node is more important.
If an operator assigns all the important nodes to the same manufacturer and it fails, network operation is greatly affected.
Therefore, the operator would assign nodes of equal importance to different manufacturers.
The same logic applies to the centrality metrics-based assignments.
The nodes are arranged in a sorted descending list according to their centrality metric values. Then, the nodes are assigned alternatively to different manufacturers.
For example, consider the topology in Fig.~\ref{fig:1} and the nodal degree (ND) centrality metric. The nodes' NDs are $\{\text{ND}(A_0) = 2, \text{ ND}(A_1) = 2, \text{ ND}(A_2) = 2, \text{ ND}(A_3) = 2, \text{ ND}(A_4) = 2, \text{ ND}(A_5) = 2, \text{{ ND}}(A_6) = 2, \text{ ND}(A_7) = 2, \text{ ND}(C_0) = 4, \text{ ND}(C_1) = 4\}$. The source and destination are disregarded in this sovereignty analysis.
The resulting sorted list in descending order of ND is $\{C_0, C_1, A_0, A_1, A_2, A_3, A_4, A_5, A_6, A_7\}$ (nodes with same ND can be written in any order). %Since this is a highly symmetric topology, several nodes have equal ND, which is uncommon in a core network.
If there are two manufacturers, the nodes $\{C_0, A_0, A_2, A_4, A_6\}$ and $\{C_1, A_1, A_3, A_5, A_7\}$ are from the first and second manufacturers respectively. 
Similarly, if there are three manufacturers, then $\{C_0, A_1, A_4, A_7\}$, $\{C_1, A_2, A_5\}$, and $\{A_0, A_3, A_6\}$ are from the first, second, and third manufacturers respectively.
Hence, the nodes of similar ND are distributed among different manufacturers.
% To make this comparison more comprehensive, three different centrality metrics are considered- Nodal Degree (ND), Betweenness Centrality (BwC), and Closeness Centrality (CC), similar to~\cite{mine-Naga, mine-ICTON-demo}. 
A comprehensive comparison is evaluated considering three different centrality metrics- Nodal Degree (ND), Betweenness Centrality (BwC), and Closeness Centrality (CC)~\cite{mine-Naga, mine-ICTON-demo}.

%A MCS-based heuristic from~\cite{mine-ICTON-demo} is also considered. 
Furthermore, manufacturer assignments from \textit{DAP}~\cite{amir}, \textit{Naga}~\cite{mine-Naga}, and \textit{Min cut set heuristic}~\cite{mine-ICTON-demo} are also compared against the \textit{CSC-ILP} to understand the improvement with \textit{CSC-ILP}.

\subsection{Evaluation Procedure}
\label{sec:evalproc}
The assignments for different topologies, numbers of manufacturers, and strategies like \textit{CSC-ILP}, \textit{DAP}, \textit{Naga}, \textit{Min cut set heuristic}, ND, BwC, and CC are obtained. First, the Network-CSC scores of all these assignments are calculated. % and plotted. 

Second, we perform all possible manufacturer failure combinations sequentially. For example, when there are three manufacturers, the possible failure combinations of the manufacturers $0$, $1$, and $2$, are $\{(0), (1), (2) (0,1), (1,2), (0,2)\}$, where, $(0,1)$ is when manufacturers $0$ and $1$ have failed simultaneously. It is not meaningful to consider $(0,1,2)$, the case where all manufacturers fail simultaneously. Then, we evaluate the Average Two Terminal Availability (ATTA). ATTA is the probability that a particular source-destination pair is connected~\cite{neumayer2010network, amir, mine-Naga}. We represent the ATTA as the percentage of successful flows under different failure scenarios. When a network is able to route all the flows, if at least any one manufacturer is operational, it can be called a fully sovereign network. However, this is unlikely in a real-world network. 
%This method to evaluate ATTA is adapted from the homogeneous analysis in Section~IV-4 in~\cite{mine1}.
%The source or destination failures are disregarded since they do not aid the sovereignty evaluation.
%This work focuses only on the connectivity of the source-destination pairs in the network.
%This work considers one flow for each possible source-destination pair in the network, except the ones with a one-hop path between the source and destination.

%\begin{center}
%\begin{figure*}[t]
%\includegraphics[width=1\linewidth, center]{New_result_images/CSC_scores_subplots.png} 
%\caption{Flow-CSC score distribution vs. Number of manufacturers ($|M|$) for different topologies. The proposed \textit{CSC-ILP} assignments outperform those based on \textit{Naga}, \textit{Min cut set heuristic}, and centrality metrics. Flow-CSC score increases with more manufacturers. The star in the box plots represents the mean. A star at the top, denoting a Flow-CSC score of 1, means that all the flows have a Flow-CSC score of 1.}
%\label{fig:three}
%\end{figure*}
%\end{center} 
\begin{figure}[t]
\includegraphics[width=1\linewidth, center]{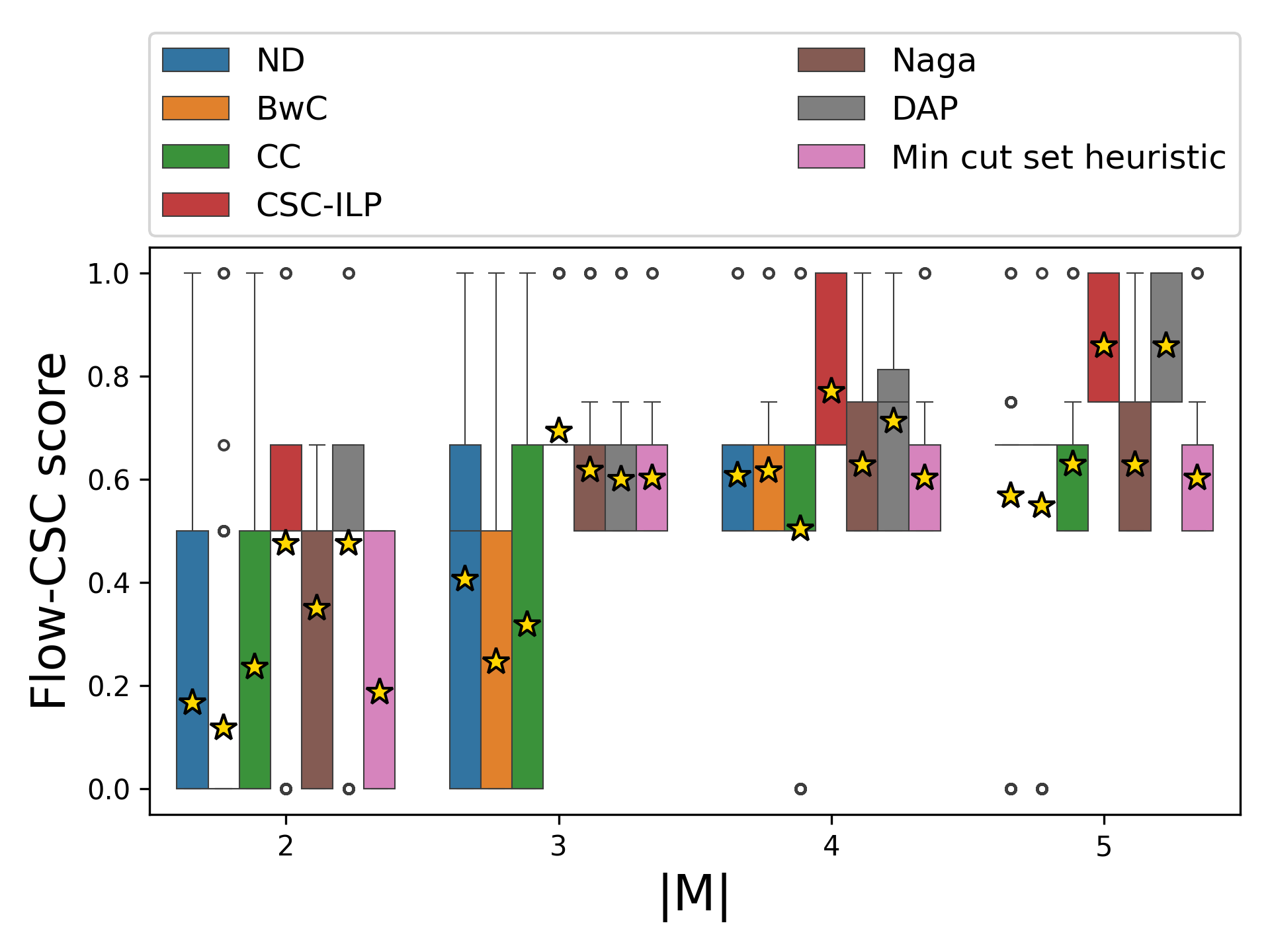} 
\caption{Network-CSC score represented as Flow-CSC score distribution vs. Number of manufacturers ($|M|$) for Polska topology. The proposed \textit{CSC-ILP} assignments outperform \textit{DAP}, \textit{Naga}, \textit{Min cut set heuristic}, and centrality metrics. Flow-CSC score increases with more manufacturers. The star in the box plots represents the mean. The heights of different color bars indicate the distribution of the central 50\% of the Flow-CSC scores. The lines on the top and the bottom of the boxes denote the remaining data range on the top and bottom quartiles, respectively. The white circles denote outliers. The absence of a box shows that the Flow-CSC scores are concentrated near the mean.}
\label{fig:three}
\end{figure}
\begin{figure}
\begin{subfigure}{1\linewidth}
    \begin{center}
    \includegraphics[width=1\linewidth]{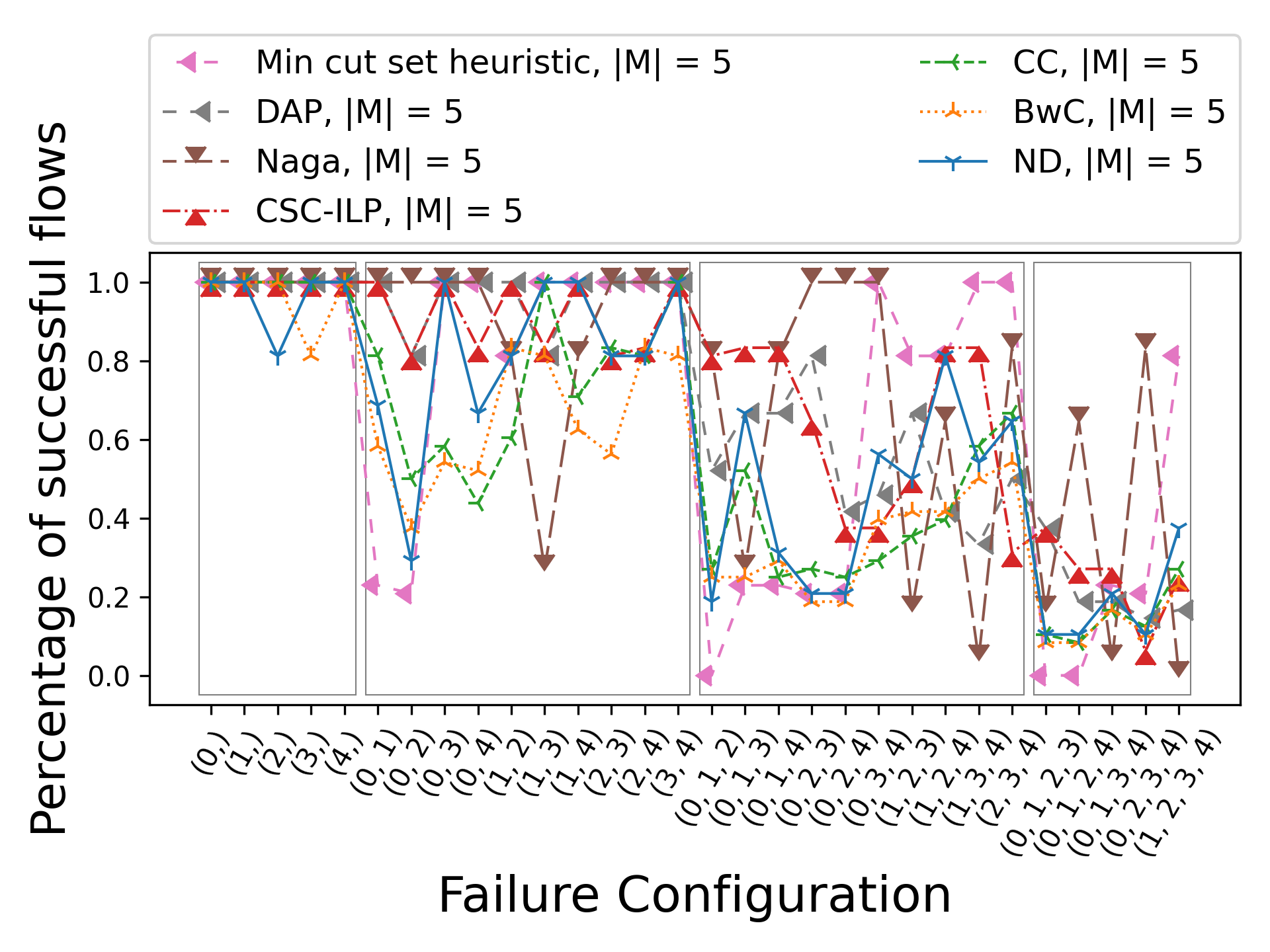}
    \end{center}
    \caption{Percentage of successful flows (ATTA) vs. Failure scenarios}  
    \label{fig:r2a}
\end{subfigure}
\begin{subfigure}{1\linewidth}
    \begin{center}
    \includegraphics[width=1\linewidth]{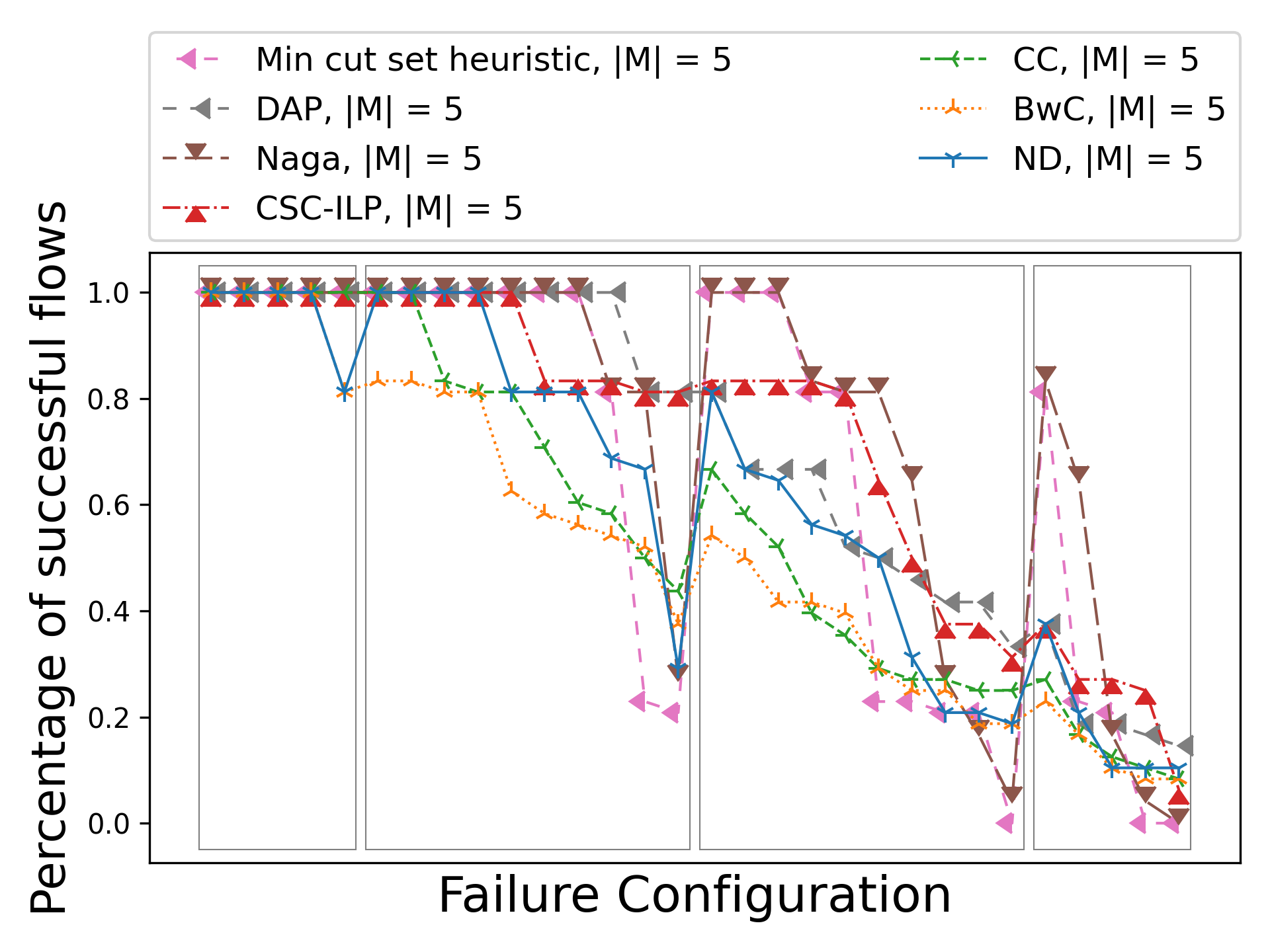}
    \end{center}
    \caption{Percentage of successful flows (ATTA) in descending order, grouped by the number of failing manufacturers}
    \label{fig:r2b}
\end{subfigure}
\caption{\textit{CSC-ILP}'s performance: Polska with five manufacturers ($|M| = 5$), \textit{CSC-ILP} outperforms other assignments.}
\label{fig:R-main}
\end{figure}
\section{Case studies- Results and Guidelines}
\label{chap:results}
\subsection{Evaluation of the CSC score and \textit{CSC-ILP}}
\label{sec:R1}
% \sha{Figures need to be updated. Explanation need to be updated with that of DAP from [New15] - new paper.}

In this section, we convey that
\begin{enumerate}[label=({V}\arabic*),align=left]
	\item\label{statement:V1} 
our proposed 
Network-CSC score is a strong indication of a network's sovereignty and
	% \item\label{statement:V2} \textit{CSC-ILP} maximizes the Network-CSC score as expected.
	\item\label{statement:V3}
\textit{CSC-ILP} outperforms the other assignments in comparison.
\end{enumerate}

%First, 
To show that the Network-CSC score is a strong indication of a network's sovereignty, we show that the successful flows under failure scenarios are the highest for \textit{CSC-ILP}. 
%For this purpose, we compare (i)~the Network-CSC score, represented as a range of all the Flow-CSC scores in a box plot for the different manufacturer assignments, as discussed in Section~\ref{sec:csc-score-description} in Fig.~\ref{fig:three}, and (ii)~the successful flow percentage under failures for the different manufacturer assignments in Fig.~\ref{fig:R-main}.
%Fig.~\ref{fig:three} shows the Flow-CSC and Network-CSC scores for manufacturer assignments 
Fig.~\ref{fig:three} shows the Network-CSC score as a box plot of all the Flow-CSC scores in the network along the Y-axis, for the manufacturer assignments from \textit{CSC-ILP}, \textit{Naga}, \textit{Min cut set heuristic}, \textit{DAP}, ND, BwC, and CC for the Polska topology.
%the Abilene, Polska, and Germany\_17 topologies. 
%The Y~and~X axes show the Flow-CSC score and different possible numbers of manufacturers ($|M|$), respectively. 
%The heights of different color bars indicate the distribution of the central 50\% of the Flow-CSC scores.
%The lines on the top and the bottom of the boxes denote the range of the remaining data on the top and bottom quartiles, respectively. The white circles on the extremes denote outliers while the star represents the mean Flow-CSC score. 
In some cases, the star denoting the mean Flow-CSC score is outside the box because the mean value is skewed by an outlier. This is why the Network-CSC score should not be represented as a single value since the distribution data can be lost.
The X-axis shows the different possible numbers of manufacturers ($|M|$) that could be available in the market for the operator.

There are two important features to note in Fig.~\ref{fig:three}- (i)~the assignment with the best average, and (ii)~the assignment with the best worst-case performance, i.e., the assignment which has the highest Flow-CSC score even in the worst case. 
Irrespective of $|M|$, the \textit{CSC-ILP} performs best in both the aforementioned features. 
The average Flow-CSC score (star) and the lowest Flow-CSC score for the \textit{CSC-ILP} for each $|M|$ is always equal or better than that of the other assignments. 
%In other words, \textit{CSC-ILP} maintains the best possible worst-case Flow-CSC scores. 
Therefore, \textit{CSC-ILP} outperforms the other manufacturer assignments and maintains the best possible worst-case Flow-CSC scores. \textit{DAP} performs close to \textit{CSC-ILP}. This is because \textit{DAP} provides the most optimal assignment to improve the average ATTA. %, as seen in Fig.~\ref{fig:R-main}.

Fig.~\ref{fig:r2a} represents ATTA by plotting the percentage of successful flows on the Y-axis against the failure combinations from Section~\ref{sec:evalproc}, along the X-axis, for the Polska topology for five manufacturers. Each line corresponds to a particular assignment. 
%The red line for \textit{CSC-ILP} stays above the other lines, showing that it outperforms the other assignments. 
To easily compare the assignments' ATTA, we plot Fig.~\ref{fig:r2b} based on Fig.~\ref{fig:r2a}. We group the failures into the grey boxes according to the number of failing manufacturers. Then, inside each group, we rearrange the lines in descending order. %Figs.~\ref{fig:r2b} shows a manufacturer assignment's relative performance against another assignment. 
%This shows the improvement in ATTA of \textit{CSC-ILP}'s assignment compared to the others.

From Fig.~\ref{fig:R-main}, ATTA decreases when more manufacturers fail. Moreover, \textit{CSC-ILP} outperforms the centrality metrics-based heuristics. At first glance into Fig.~\ref{fig:r2b}, \textit{DAP}, \textit{Min cut set heuristic}, and \textit{Naga} seem to outperform \textit{CSC-ILP} in some cases. However, \textit{CSC-ILP} maintains the best possible worst-case ATTA. For example, for the two-manufacturer failure scenario (grey box in the middle) in Fig.~\ref{fig:r2b}, the worst possible ATTA for \textit{CSC-ILP} and \textit{Naga} is 0.45 and 0.25 respectively. Therefore, \textit{CSC-ILP} minimizes the impact of the worst-case manufacturer failures as expected. \textit{DAP} provides strong ATTA results because it is globally optimized to maximize average ATTA in the network. However, it has weaker worst-case performance and is computationally very intensive. 
%For example, \textit{DAP} took 5 hours for Nobel\_EU for four manufacturers, while \textit{CSC-ILP} took 1.5 minutes. 
\textit{DAP}'s scalability issue is addressed in~\cite{amir} and the authors propose heuristics to be feasible. 

The results observed are consistent for all the sample topologies.
Therefore, the graphs shown in Figs.~\ref{fig:three} and~\ref{fig:R-main} support the statements~\ref{statement:V1}~and~\ref{statement:V3}. 
Now, we can proceed with our Network-CSC score analysis.

\subsection{Analysis and Inferences from the Network-CSC scores}
\label{sec:R2}
This section answers important questions from an operator's standpoint.

\subsubsection{What is the impact of purchasing components from more manufacturers?} \hfill\\
Fig.~\ref{fig:three} shows that more manufacturers increase the Flow-CSC scores because when more manufacturers are available, the dependency on one manufacturer is reduced. Therefore, the network sovereignty increases. 
Using \textit{CSC-ILP} with more manufacturers is more beneficial to the operator.

\subsubsection{How many manufacturers does my network require?} \hfill\\
%Abilene, in Fig.~\ref{fig:three}a, is a small topology with an average MCS cardinality of 2. The Network-CSC score saturates at one when there are three or more manufacturers.
%For Polska in Fig.~\ref{fig:three}, with an average MCS cardinality of around 3.2, the average of the Flow-CSC scores nearly reaches one for five manufacturers. 
%For Germany\_17 in Fig.~\ref{fig:three}c, with an average MCS cardinality of around 2.5, the average of the Flow-CSC scores nearly reaches one for four manufacturers.
%These examples show that the number of manufacturers must be greater than the average MCS cardinality for a strong Network-CSC score.
The ideal number of manufacturers required for a network depends on the number of MCSs for the flows and the cardinality of the MCSs. Ideally, fewer MCSs with high cardinality are a sign of strong robustness. This is because fewer MCSs mean fewer possibilities for the flow to fail, and high MCS cardinality means several nodes must fail for that MCS to fail. The probability that all the elements of a large MCS fail simultaneously is minimal. 

Though the inherent structural robustness of the topology is beyond the network operator's control in most cases, the operator can still improve network sovereignty by having maximum manufacturer diversity, even in the larger MCSs. For example, the average Flow-CSC score for Polska in Fig.~\ref{fig:three} is nearly 1.0 for five manufacturers ($|M| = 5$). This is because the average MCS cardinality for Polska is 3.2, with ten MCSs in the network for different flows having a cardinality of 4. Hence, it is advisable to have more manufacturers in the network than the highest MCS cardinality in the network. This inference also agrees with the results in the other topologies.

%From these examples, we can deduce that the number of manufacturers required to guarantee a strong Network-CSC score must be greater than the average MCS cardinality.
% \sha{to do for Shakthi: Calculate average MCS cardinality properly.}
% In this case, the Network-CSC score can have only two values. If the same manufacturer is used in the MCS, the MCS-CSC score is zero. If two different manufacturers are used in the MCS, then the MCS-CSC score is one. Therefore, the box plot is divided between zeroes and ones. However, having more ones is better because such an assignment avoids manufacturer dependency as much as possible. 

\subsubsection{Why do some flows have a Flow-CSC score equal to 0?}\hfill\\
%Some flows have Flow-CSC score equal to 0 even for \textit{CSC-ILP}, as seen in Fig.~\ref{fig:three}c for Germany\_17 for two manufacturers. This occurs because some flows have at least one low cardinality MCS with nodes from the same manufacturer. Therefore, \textit{CSC-ILP} prioritizes the other Flow-CSC scores to improve the Network-CSC score. However, this behavior is unavoidable and is also observed in other assignments.
Some flows have Flow-CSC scores of 0. For example, CC, \textit{Naga}, and the \textit{Min cut set heuristic} for two manufacturers ($|M| = 2$) have Flow-CSC score of 0, shown by the outliers in Fig.~\ref{fig:three}. This occurs because some flows have at least one low cardinality MCS with nodes from the same manufacturer.% Therefore, \textit{CSC-ILP} prioritizes the other Flow-CSC scores to improve the Network-CSC score.
However, this behavior is unavoidable when the number of manufacturers is low and the assignment is sub-optimal.

\subsubsection{How should the network operator map the real manufacturers to the manufacturer IDs used in this study?}\hfill\\
Fig.~\ref{fig:r2a} shows the manufacturer(s) causing the most impactful failures.
For example, for \textit{CSC-ILP}, for the Polska topology, for four manufacturers, a combination of manufacturers $M_1$ and $M_3$ has the highest impact on failing, shown by the low Flow-CSC score of 0.5 for the red triangle in the middle grey box.
Therefore, the operator must assign the most trustworthy manufacturers to the manufacturer IDs $M_1$ and $M_3$.
Since there is no direct mapping between manufacturers in the physical world and the manufacturer-IDs used in this study, this is up to the operator to decide. 
Hence, this analysis helps the operator avoid the worst-case scenarios possible.

%\subsection{Comparison with previous works}
%
%Consider the example DCN topology in Fig.~\ref{fig:1}. Consider four different manufacturers (yellow, green, blue, and red). Applying \textit{CSC-ILP} to this topology, the manufacturer assignment obtained is the same as the example assignment discussed in Section~\ref{sec:example-applicaiton}, also shown as the colors of the nodes in Fig.\ref{fig:1}. This assignment maximizes the diversity inside each MCS, preventing a single point of failure. It ensures the flow's survival even if one or more manufacturers fail, provided both core switches are available. This agrees with the guideline in Section\ref{sec:guideline_for_sov}. 
%
%Our previous work~\cite{mine2} on DCN sovereignty showed that a linearly left-right coloring scheme (`Left-right sequential' arrangement, similar to the result we obtained with \textit{CSC-ILP}, was the most sovereign heuristic manufacturer assignment possible. This further validates the CSC-score metric as a good measure of network sovereignty.

\subsection{Guidelines to operators}
\label{sec:guidelines}
Based on these results, we provide the following guidelines to operators.
\begin{enumerate}[label=({G}\arabic*),align=left]
    \item It is unsafe to purchase most components from one manufacturer. A fairly distributed manufacturer assignment is advisable.
    \item The number of manufacturers required depends on the average MCS cardinality. If the number of manufacturers exceeds the average MCS cardinality, the network is more sovereign.
    \item Having several manufacturers might be difficult in a real-life scenario due to a limited number of manufacturers in the market and interoperability issues.
    \item Manufacturer dependency can be avoided by using the CSC score and \textit{CSC-ILP} for manufacturer assignment.
\end{enumerate}

\subsection{Network-CSC score and \textit{CSC-ILP} Limitations}
There are several networking devices from different manufacturers that use a particular subcomponent from the same subcomponent manufacturer.
Hence, the Network-CSC score and \textit{CSC-ILP}, currently focusing on component-level sovereignty, need to be further adapted to consider subcomponent-sharing.
Moreover, the Network-CSC score does not consider the probability of a particular manufacturer failing. Modeling the trustworthiness of a manufacturer remains an open challenge.
Additionally, the \textit{CSC-ILP} must be extended to optimize for other dependability attributes discussed in~\cite{Avizienis:2004}. 

% \begin{figure*}
% \begin{subfigure}{0.33\linewidth}
%     \begin{center}
%     \includegraphics[width=1\linewidth]{New_result_images/Germany_17_CSC_score_M_comparison_Metric.png}
%     \end{center}
%     \caption{Germany\_17 topology} 
%     \label{fig:r1}
% \end{subfigure}
% \begin{subfigure}{0.33\linewidth}
%     \begin{center}
%     \includegraphics[width=1\linewidth]{New_result_images/Abilene_CSC_score_M_comparison_Metric.png}
%     \end{center}
%     \caption{Abilene topology}  
%     \label{fig:abilene}
% \end{subfigure}
% \begin{subfigure}{0.33\linewidth}
%     \begin{center}
%     \includegraphics[width=1\linewidth]{New_result_images/Polska_CSC_score_M_comparison_Metric.png}
%     \end{center}
%     \caption{Polska topology}
%     \label{fig:polska}
% \end{subfigure}
% \caption{Flow-CSC score distribution vs. Number of manufacturers, \textit{CSC-ILP} outperforms centrality metrics, Flow-CSC score increases with more manufacturers, and having a number of manufacturers greater than or equal to the average MCS cardinality maximizes the Network-CSC score.}
% %\caption{CSC score per flow vs. Number of manufacturers, \textit{CSC-ILP} outperforms centrality metrics, Network-CSC score increases with more manufacturers, and having a number of manufacturers greater than or equal to the average MCS cardinality maximizes the Network-CSC score.}
% \label{fig:three}
% \end{figure*}
\squeezeup
\section{Conclusion and Outlook}
\label{chap:conclusion}
This work introduced the novel CSC score metric to quantify and compare the network sovereignty of different manufacturer assignments and networks.
%, to guide network operators to make well-informed design choices.
We proposed the \textit{CSC-ILP} to minimize the impact of manufacturer failures in networks by removing dependencies on the manufacturer(s). The results show the CSC score's validity as a sovereignty metric and \textit{CSC-ILP}'s solution as a fast and scalable solution to maximize network sovereignty.
%We also showed the improved performance of the \textit{CSC-ILP} compared to the other works in the state of the art.
In the future, we aim to combine sovereignty and reliability for robust network designs.
%Through this work, we hope to trigger fellow researchers to indulge in network sovereignty studies. 

\bibliographystyle{IEEEtran}
\bibliography{IEEEabrv, 2_Bibliography}
%\newpage

\section{Biography Section}


\vspace{-15pt} \squeezeup \squeezeup 
\begin{IEEEbiographynophoto}{Shakthivelu Janardhanan} joined the Chair of Communication Networks, Technical University of Munich (TUM) as a doctoral candidate in 2021 focusing on network dependability.
\end{IEEEbiographynophoto}
\squeezeup \squeezeup  \squeezeup \squeezeup 
\begin{IEEEbiographynophoto}{Ritanshi Agarwal} joined the Chair of Communication Networks at Universität der Bundeswehr (UniBw) München as a research associate in 2024, focusing on resilient optical access networks.
\end{IEEEbiographynophoto}
\squeezeup \squeezeup \squeezeup \squeezeup 
\begin{IEEEbiographynophoto}{Wolfgang Kellerer} 
is a Full Professor at TUM, heading the Chair of Communication Networks. 
%Before, he was for over ten years with NTT DOCOMO’s European Research Laboratories. 
He's an editor for IEEE Transactions on Network and Service Management and Network Virtualization for IEEE Communications Surveys and Tutorials.
\end{IEEEbiographynophoto} 
\squeezeup \squeezeup \squeezeup \squeezeup 
\begin{IEEEbiographynophoto}{Carmen Mas-Machuca}
is a Full professor at UniBw, heading the Chair of Communication Networks.
%Her research interests include techno-economic studies, resilient network planning, and optical networks. 
She's an editor for IEEE Transactions on Network and Service Management, OSA Journal on Optical Communications and Networking, and IEEE Communications Magazine.
%and is also active in international conferences as chair, TPC co-chair, and TPC member.
\end{IEEEbiographynophoto}
\vspace{11pt}

%\newpage 
%
%\ % The empty page
%
%\newpage
%
%\begin{equation*}
%A_f = A_{T_0} \times \bigl(1-\prod\limits_{i=0}^{3}(1 - A_{A_i})\bigr) \times \bigl(1-\prod\limits_{i=0}^{1}(1 - A_{C_i})\bigr) \times \bigl(1-\prod\limits_{i=4}^{7}(1 - A_{A_i})\bigr) \times A_{T_1}
%\end{equation*}

%\begin{equation*}
%\color{red} A_{0||1} = 1 - (1-A_{A_0})(1-A_{A_1})
%\end{equation*}
%
%\begin{equation*}
%\color{blue} A_{2||3} = 1 - (1-A_{A_2})(1-A_{A_3})
%\end{equation*}
%
%\begin{equation*}
%\color{orange} A_{4||5} = 1 - (1-A_{A_4})(1-A_{A_5})
%\end{equation*}
%
%\begin{equation*}
%\color{violet}A_{6||7} = 1 - (1-A_{A_6})(1-A_{A_7})
%\end{equation*}
%
%\begin{equation*}
%A_f = A_{T_0} \times \biggl[1- \Bigl(1-\bigl(A_{0||1} \cdot A_{C_0} \cdot A_{4||5}\bigr)\Bigr)\Bigl(1-\bigl(A_{2||3} \cdot A_{C_1} \cdot A_{6||7}\bigr)\Bigr)\biggr] \times A_{T_1}
%\end{equation*}

%\begin{equation*}
%A_{x||y} = 1 - (1-A_{A_x})(1-A_{A_y})
%\end{equation*}

\end{document}